\documentclass[a4paper]{jpconf}
\usepackage{graphicx}
\usepackage{epsfig}
\begin{document}
\def\b{\bar}
\def\d{\partial}
\def\D{\Delta}
\def\cD{{\cal D}}
\def\cK{{\cal K}}
\def\f{\varphi}
\def\g{\gamma}
\def\G{\Gamma}
\def\l{\lambda}
\def\L{\Lambda}
\def\M{{\Cal M}}
\def\m{\mu}
\def\n{\nu}
\def\p{\psi}
\def\q{\b q}
\def\r{\rho}
\def\t{\tau}
\def\x{\phi}
\def\X{\~\xi}
\def\~{\widetilde}
\def\h{\eta}
\def\bZ{\bar Z}
\def\cY{\bar Y}
\def\bY3{\bar Y_{,3}}
\def\Y3{Y_{,3}}
\def\z{\zeta}
\def\Z{{\b\zeta}}
\def\Y{{\bar Y}}
\def\cZ{{\bar Z}}
\def\`{\dot}
\def\be{\begin{equation}}
\def\ee{\end{equation}}
\def\bea{\begin{eqnarray}}
\def\eea{\end{eqnarray}}
\def\half{\frac{1}{2}}
\def\fn{\footnote}
\def\bh{black hole \ }
\def\cL{{\cal L}}
\def\cH{{\cal H}}
\def\cF{{\cal F}}
\def\cP{{\cal P}}
\def\cM{{\cal M}}
\def\ik{ik}
\def\mn{{\mu\nu}}
\def\a{\alpha}

\title{Gravity vs. Quantum theory:  Is electron really  pointlike?}

\author{Alexander Burinskii}

\address{Lab. of Theor. Phys. , NSI, Russian Academy of
Sciences, B. Tulskaya 52  Moscow 115191 Russia}

\ead{bur@ibrae.ac.ru}

\begin{abstract}
The observable gravitational and electromagnetic parameters of an
electron: mass $m$, spin $J=\hbar/2$, charge $e$ and magnetic
moment $ea = e\hbar /(2m)$ indicate unambiguously that the
electron should had the Kerr-Newman background geometry -- exact
solution of the Einstein-Maxwell gravity for a charged and
rotating black hole. Contrary to the widespread opinion that
gravity plays essential role only on the Planck scales, the
Kerr-Newman gravity displays a new dimensional parameter $a
=\hbar/(2m),$ which for parameters of an electron corresponds to
the Compton wavelength and turns out to be very far from the
Planck scale.  Extremely large spin of the electron with respect
to its mass produces the Kerr geometry without horizon, which
displays very essential topological changes at the Compton
distance resulting in a two-fold structure of the electron
background. The corresponding gravitational and electromagnetic
fields of the electron are concentrated near the Kerr ring,
forming a sort of a closed string, structure of which is close to
the described by Sen heterotic string. The indicated by Gravity
stringlike structure of the electron contradicts to the statements
of Quantum theory that electron is pointlike and structureless.
However, it confirms  the peculiar role of the Compton zone of the
"dressed" electron and matches with the known limit of the
localization of the Dirac electron.
 We discuss the relation of the Kerr string with the low energy string theory
 and with the Dirac theory of electron and suggest that the predicted by the
 Kerr-Newman gravity closed string in the core of the electron, should be experimentally observable by
 the novel regime of the high energy scattering -- the
  Deeply Virtual (or "nonforward") Compton Scattering".
\end{abstract}

\section{Introduction}
Modern physics is based on Quantum theory and Gravity. The both
theories are confirmed experimentally with great precision.
Nevertheless, they are conflicting and cannot be unified in a
whole theory. General covariance is main merit of General
Relativity and the main reason of misinterpretation. The freedom
of coordinate transformations is one of the source of the
conflict. One of the source of the problems is the absence of the
usual plane waves in general relativity, which causes the conflict
with the Fourier transform and prevents expansion of the quantum
methods to the curved spacetimes.

The analogs of the plane waves in gravity are the pp-wave
solutions which are singular, either at infinity or at some
lightlike (twistor) line, forming a singular ray similar to the
laser beam \cite{HorSt}. The pp-wave singular beams are modulated
by the usual plane waves and form singular strings, which are in
fact the fundamental strings of the low energy string theory.
Moreover, it turns out that the pp-waves don't admit $\alpha '$
stringy corrections \cite{HorSt,Tseyt,Coley}, and therefore they
are exact solutions to the full string theory.

The null Killing
 direction of the pp-waves,  $k_\m , \quad (k_\m k^\m = 0 ) ,$  is  adapted to the
 Kerr-Schild (KS) form of the metric  $g^\mn
=\eta^\mn + 2H k^\m k^\n $ which is rigidly related with the
auxiliary Minkowski background geometry $\eta^\mn$. The KS class
of metrics is matched with the light-cone structure of the
Minkowski background which softens conflict with quantum theory.
In spice of
 the extreme rigidity, the Kerr-Schild coordinate system allows one to
 describe practically all the physically interesting
 solutions of General Relativity, for example:
 \begin{itemize}

\item rotating black holes and the sources without horizon,

\item de Sitter and Anti de Sitter spaces, and their rotating
analogues,

\item combinations of a black hole inside the de Sitter or AdS
 background spacetime,

\item the opposite combinations: dS or AdS spaces as regulators of
the black hole geometry

\item charged black holes and rotating stars, and so on.

\end{itemize}

In particular, the Kerr-Schild metric describes the Kerr-Newman
(KN) solution for a charged and rotating black hole, which for the
case of very large angular momentum may be considered as a model
of spinning particles.

In  1968 Carter obtained,  \cite{Car,DKS}, that the KN solution
has the gyromagnetic ratio $g = 2$ as that of the
 Dirac electron, which initiated a series of the works on the
KN electron model
\cite{Car,DKS,BurSol,BurTwi,BurKN,Isr,BurGeonIII,Bur0,IvBur,Lop,BurSen,BurStr,
BurBag,Arc,Dym,TN}.

 In this
paper we discuss  one of the principal contradictions between
Quantum theory and Gravity, the question on the shape and size of
the electron,  believing that the nature of the electron is
principal point for understanding of Quantum Theory. Quantum
theory states that electron is pointlike and structureless. In
particular,
 Frank Wilczek  writes in \cite{FWil}: "...There's no evidence that electrons have
internal structure (and a lot of evidence against it)", while the
superstring theorist Leonard Susskind notes
 that  electron radius is "...most probably not much bigger and
not much smaller than the Planck length..",   \cite{LSuss}.
 This point of view is supported by the experiments
with high energy scattering, which have not found the electron
structure down to $10^{-16} cm .$

On the other hand, the experimentally observable parameters of the
electron: angular momentum $J$, mass $m$, charge $e$ and the
magnetic moment $ \mu $ indicate unambiguously that its background
geometry should be very close to the corresponding Kerr-Newman
(KN) solution of the Einstein-Maxwell field equations. The
observed parameters of the electron $J,m,e, \mu $ determine also
the corresponding parameters of the KN solution: the mass $m ,$
charge $e ,$ and the new dimensional parameter $a =L/m $ which
fixes radius of the Kerr singular ring. The fourth observable
parameter of the electron, magnetic momentum $\mu ,$ conform to
the KN solution automatically as consequence of the above
discussed specific gyromagnetic ratio of the Dirac electron
coinciding with that of the KN solution. As a result, the KN
solution indicates a characteristic radius of the electron as the
Compton one $a=\hbar/(2m) ,$ corresponding to the radius of the
Kerr singular ring. Therefore, contrary to Quantum theory, the KN
gravity predicts the ring like structure of the electron and its
Compton size.

The metric of the Kerr-Newman spacetime has the form
 \be g_\mn=\eta _\mn + 2 H k_\m k_\n , \label{ksm} \ee
 where \be H=\frac {mr - e^2/2}{r^2 + a^2 \cos ^2
\theta} , \label{H} \ee and the electromagnetic (EM) vector
potential of the KN solution is \be \alpha^\m_{KN} = Re \frac e
{r+ia \cos \theta} k^\m  \label{ksGA}, \ee where $r$ and $\theta$
are Kerr's oblate spheroidal coordinates which are related to the
Cartesian coordinates as follows

\bea \nonumber x+iy  &=& (r + ia) e^{i\phi} \sin \theta  \\
z &=&r\cos\theta . \label{oblate}\eea

 The
metric and EM field are aligned with the null vector field $k^\m$
forming a Principal Null Congruence (the `Kerr congruence'), see
Figure.1. The Kerr congruence is determined by
 {\it the Kerr Theorem} in  twistor terms (each line of the congruence
 is a twistor null line). Although the KN spacetime is curved,
 the Kerr congruence foliates it into a family of the flat complex
 twistor null planes, which allows one to use in the Kerr-Schild spaces
 a twistor version of the Fourier transform, which forms a holographic bridge between
the classical Kerr-Schild gravity and
 Quantum theory, \cite{BurExa,BurPreQ}.

\begin{figure}[h]
\begin{minipage}{17pc}
\includegraphics[width=17pc]{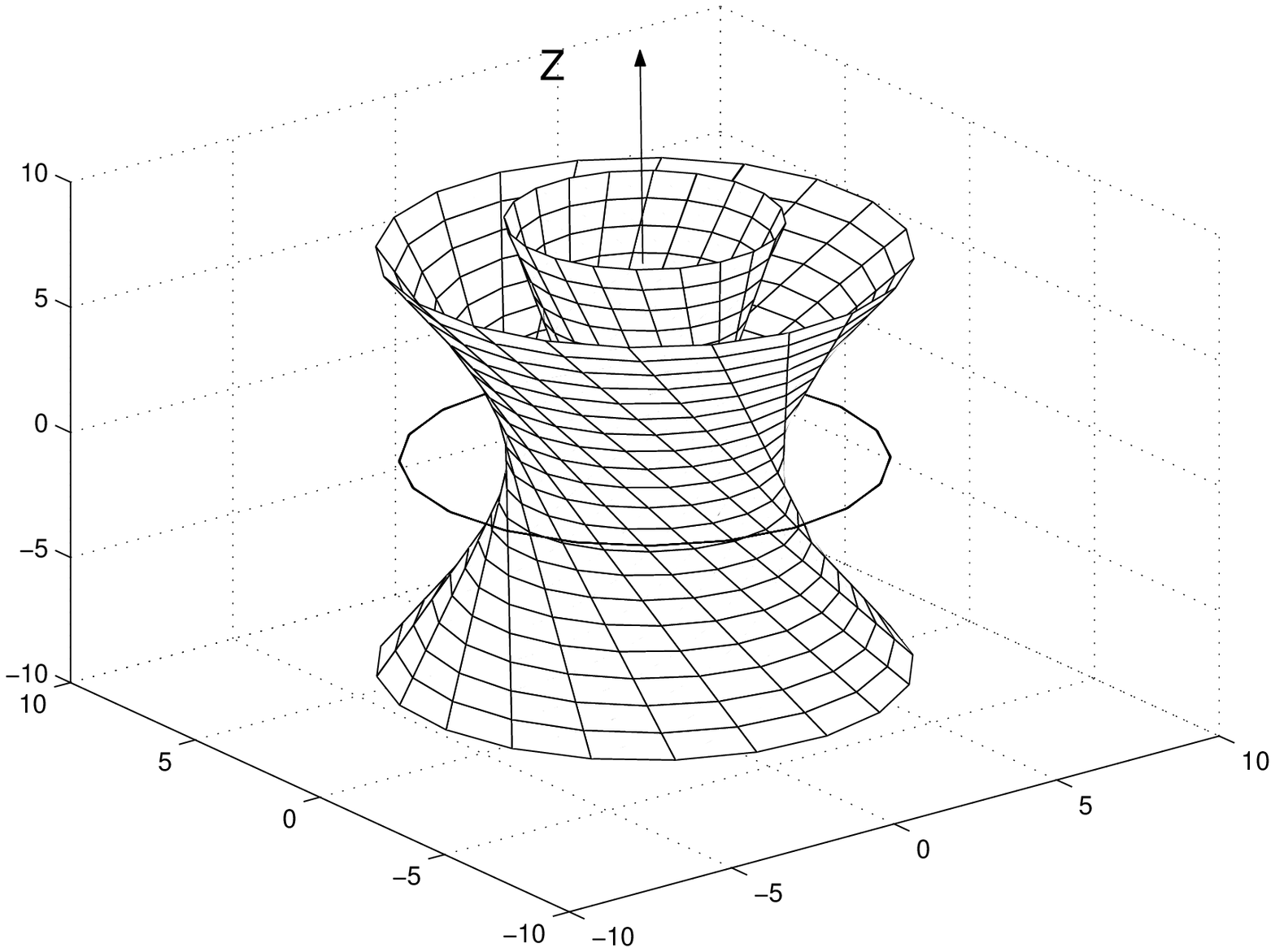}
\caption{\label{label}Vortex of the Kerr congruence. Twistor null
lines are focused on the Kerr singular ring, forming a circular
gravitational waveguide, or string with lightlike excitations.}
\end{minipage}\hspace{6pc}%
\begin{minipage}{14pc}
\includegraphics[width=15pc]{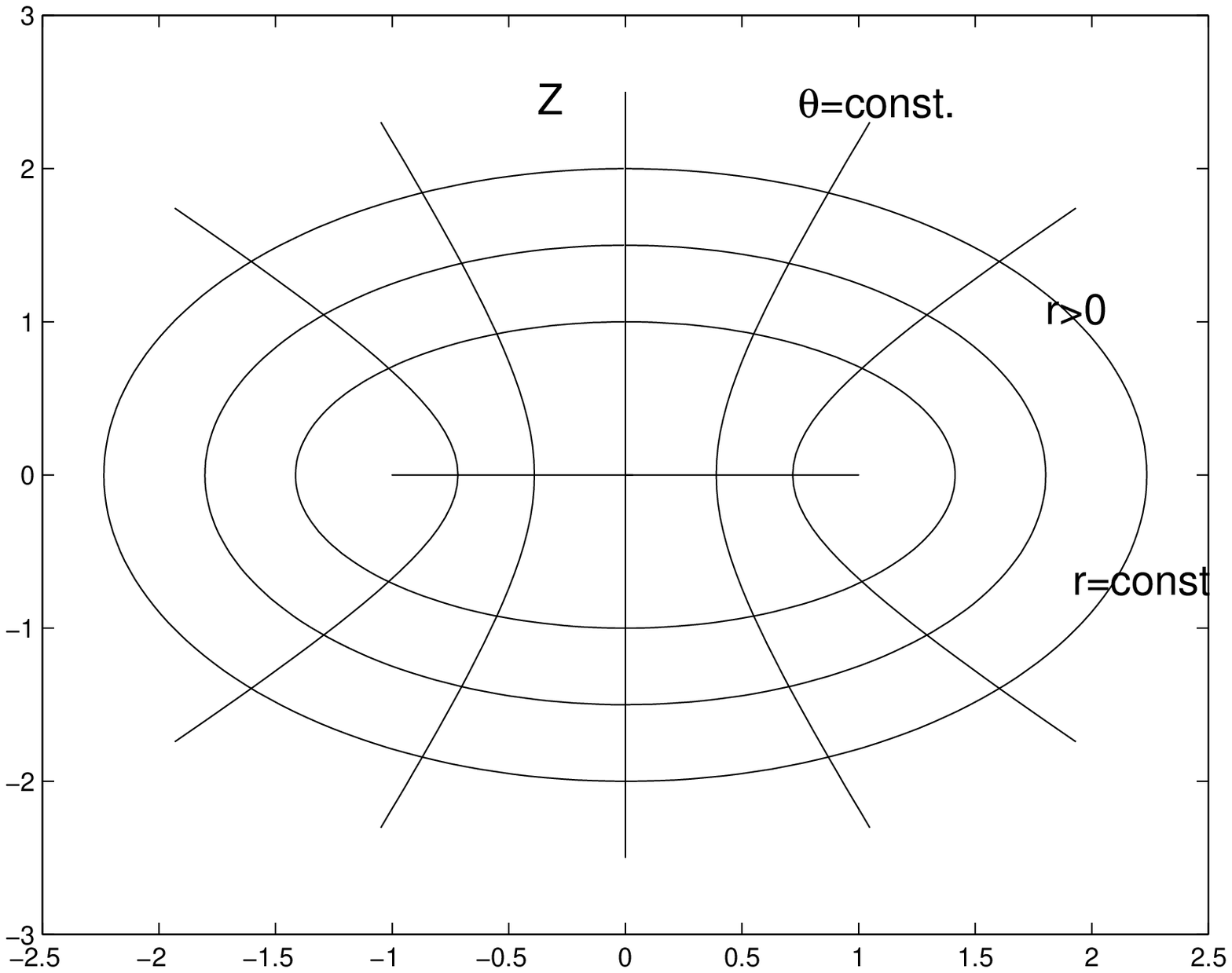}
\caption{\label{label}Oblate coordinate system $r, \ \theta $ with
focal points at $r=\cos\theta = 0$ forms a twofold analytic
covering: for $r>0$ and $r<0.$ }
\end{minipage}
\end{figure}

The KN gravitational and EM fields are concentrated near the Kerr
singular ring, which appears in the rotating BH solutions instead
of the pointlike Schwarzschild singularity. One sees that
 radius of the ring $a =J/m$ increases for the small masses and is proportional to
the spin $J.$ Therefore, contrary to the characteristic radius of
the Schwarzschild solution (related with position of the BH
horizon, $ r_g=2m $), the characteristic extension of the KN
gravitational field turns out to be much beyond the Planck length,
and corresponds to the Compton length, $r_{compt}= a =\hbar/(2m),
$ or to the radius of a "dressed" electron. In the units $c=\hbar
=G=1 , $ mass of the electron is $m\approx 10^{-22},$ while $ \
a=J/m \approx 10^{22} .$ Therefore, $a>>m ,$ and the black hole
horizons disappear, showing that the Kerr singular ring is naked.
In this case the Kerr spacetime turns out to be twosheeted, since
the Kerr ring forms its branch line creating a twosheeted
topology. The relations (\ref{ksm}) and (\ref{ksGA}) show that the
gravitational and electromagnetic fields of the KN solution are
concentrated in a thin vicinity of the Kerr singular ring
$r=\cos\theta=0,$ forming a type of ``gravitational
 waveguide'', or a closed string, \cite{IvBur}.
The Kerr string takes the Compton radius, corresponding to the
size of a "dressed" electron in QED and to the limit of
localization of the electron in the Dirac theory \cite{BjoDr}.

There appear two questions:

(A) How does the KN gravity know about one of the principal
parameters of Quantum theory? and

(B) Why does Quantum theory works successfully on the flat
spacetime, ignoring the stringy defect of the background geometry?

A small and slowly varying gravitational field could be neglected,
however
 the stringlike KN singularity forms a branch-line  of the
KS spacetime, and such a topological defect cannot be ignored. A
natural resolution of this trouble could be the assumption that
there is an underlying theory providing the consistency of quantum
theory and gravity.
 In this paper we suggest a  rather unexpected resolution of this
 puzzle, claiming that underlying theory is the low energy string theory,
 in which the closed string is created by the KN gravity related with
 twistorial structure of the Kerr-Schild pre-quantum geometry \cite{BurExa,BurPreQ}.
 The Kerr singular ring is generated as a
caustic of the Kerr twistor congruence and  forms a closed string
on the boundary of the Compton area of the electron. The KN
gravity indicates  that this string should represent a principal
element of the extended electron structure.

If the closed Kerr string is really formed on the boundary of the
Compton area, it should be experimentally observable. There
appears the question while it was not obtained earlier in the high
energy scattering experiments. We find some explanation to this
fact and arrive at the conclusion that the KN string should
apparently be detected by the novel experimental regime of the
high energy scattering which is based on the theory of Generalized
Parton Distributions (GPD), and corresponds to a ``non-forward
Compton scattering'' \cite{Rad,Ji}, suggested recently for
tomography of the particle images \cite{Hoyer}.

\section{Twosheetedness of the Kerr-Geometry}

In the KS representation \cite{DKS}, a few coordinate systems are
used simultaneously. In particular, {\it the null Cartesian
coordinates}
\[ \z = (x+iy)/\sqrt 2 , \
 \Z = (x-iy)/\sqrt 2 , \ u = (z - t)/\sqrt 2 , \
v = (z + t)/\sqrt 2 \] are used for description of the Kerr
congruence in the differential form  \be k_\m dx^\m = P^{-1}( du +
\bar Y d \zeta + Y d \bar\zeta - Y \bar Y dv), \label{kY}\ee via
the complex function $Y(x)=e^{i\phi} \tan \frac \theta 2 ,$ which
is a projective angular coordinate on the celestial sphere, \be
Y(x)=e^{i\phi} \tan \frac \theta 2 . \label{Y} \ee

\subsection{The Kerr Theorem}
 Kerr congruence (PNC) is controlled by {\it
THE KERR THEOREM:}

The geodesic and shear-free Principal null congruences (type D
metrics) are determined by  holomorphic function $Y(x)$ which is
analytic solution of the equation \be F (T^a) = 0 \ , \ee where
$F$ is an arbitrary analytic function of the projective twistor
coordinates \be T^a =\{ Y,\quad \z - Y v, \quad u + Y \Z \} .\ee

The Kerr theorem is a practical tool for the obtaining the exact
Kerr-Schild solutions. The following sequence of steps is assumed:
\[ F (T^a) =0 \Rightarrow F (Y, x^\m) = 0 \Rightarrow Y(x^\m)
\Rightarrow k^\m (x) \Rightarrow g^\mn \] For the Kerr-Newman
solution function $F$ is quadratic in $Y ,$ which yields TWO roots
$Y^\pm(x)$ corresponding to two congruences!

 As a result the obtained two  congruences (IN and OUT) determine two sheets of
 the Kerr solution: the ``negative (--)" and
``positive (+)" sheet, where the fields change their directions.
In particular, two different congruences $ k^{\m(+)} \ne
k^{\m(-)}$ determine two different KS metrics $ g_\mn^{(+)} \ne
g_\mn^{(-)} $ on the same Minkowski background. As it shows the
Fig.1, the Kerr congruence propagates analytically from IN to OUT-
sheet via the disk $r=0 ,$  and therefore, the two KS sheets are
linked analytically. The twosheeted KN space is parametrized by
the oblate spheroidal coordinate system $r, \ \theta, \phi ,$
which tends asymptotically, by $r\to \infty ,$ to the usual
spherical coordinate system.  Twosheetedness is the long-term
mystery of the Kerr solution!
 For the
multiparticle Kerr-Schild (KS) solutions, \cite{Multiks}, the Kerr
theorem yields many roots $Y^i, \ i=1,2,...$ of the Kerr equation
$F(Y)=0 ,$ and the KS geometry turns out to be {\it multivalued}
and {\it multisheeted}.

 The extremely simple form of the
Kerr-Schild metric (\ref{ksm}) is related with complicate form of
the Kerr congruence, which represents a type of deformed (twisted)
hedgehog. In the rotating BH solutions the usual pointlike
singularity inside the BH turns into a \emph{a closed singular
ring,} which is interpreted as a closed string in the
corresponding models of the spinning elementary particles
\cite{BurStr}. The KN twosheetedness was principal puzzle of the
Kerr geometry for four decades and determined development of the
KN electron models along two principal lines of investigation: I)
the bubble models, and II) the stringlike models.

I. -- In 1968 Israel suggested to  truncate negative KN sheet,
$r<0 ,$, and replace it by the {\it rotating disklike source} (
$r=0 $) spanned by the Kerr singular ring of the Compton radius
$a=\hbar/2m ,$ \cite{Isr}. Then, Hamity obtained in \cite{Ham}
that the disk has to be rigidly rotating, which led to a
reasonable interpretation of the matter of the source as an exotic
stuff with zero energy density and negative pressure. The matter
distribution appeared singular at the disk boundary, forming an
additional closed string source, and L\'opez suggested in
\cite{Lop} to regularize this source, covering the Kerr singular
ring by a disklike ellipsoidal surface. As a result, the KN source
was turned into a rotating and charged oblate bubble with a flat
interior, and further it was realized as a regular soliton-like
bubble model \cite{BurSol}, in which the boundary of the bubble is
formed by a domain wall interpolating between the external KN
solution and a flat pseudovacuum state inside the bubble.

II. The stringlike models of the KN source retain the  twosheeted
topology of the KN solution, forming a closed 'Alice' string of
the Compton size \cite{IvBur,BurKN,BurAxi}. The Kerr singular ring
is considered as a waveguide for electromagnetic traveling waves
generating the spin and mass of the KN solution in accordance with
the old Wheeler's "geon" model of `mass without mass'
\cite{Wheel,BurGeonIII,Bur0,BurGeon0}.\fn{Note also that the KN
twosheetedness represents a natural realization of the
Einstein-Rosen bridge model related with the ``in-going`` and
``out-going'' radiation \cite{BurA}, as well as with the Wheeler
``charge without charge'' model.} In this paper we concentrate on
the stringlike model of the electron, which displays close
relations to the low energy string theory. The bubble model of
regularization of the KN solution is discussed briefly in sec.5
along our previous papers.

\section{The Kerr singular ring as a closed string}

Exact {\it non-stationary} solutions for electromagnetic
excitations on the Kerr-Schild background,
\cite{BurAxi,BurA,BurExa}, showed that there are no smooth
harmonic solutions. The typical exact electromagnetic solutions on
the KN background take the form of singular beams propagating
along the rays of PNC, contrary to smooth angular dependence of
the wave solutions used in perturbative approach!

 Position of the horizon for the excited KS  black holes solutions is determined by
function $H $ which has for the exact KS solutions the form,
\cite{DKS}, \be H =\frac {mr - |\psi|^2/2} {r^2+ a^2 \cos^2\theta}
 \ , \label{Hpsi} \ee where $\psi(x)$ is related to the vector potential
of the electromagnetic field
\be \alpha =\alpha _\m dx^\m \\
= -\frac 12 Re \ [(\frac \psi {r+ia \cos \theta}) e^3 + \chi d \Y
], \quad \chi = 2\int (1+Y\Y)^{-2} \psi dY  \ , \label{alpha} \ee
 which obeys
the alignment condition \be \alpha _\m k^\m=0 . \ee The equations
(\ref{ksm})and (\ref{Hpsi}) display compliance and elasticity of
the horizon with respect to the electromagnetic field.

The Kerr-Newman solution corresponds to $\psi=q=const.$. However,
any nonconstant holomorphic function $\psi(Y) $ yields also an
exact KS solution, \cite{DKS}. On the other hand, any nonconstant
holomorphic functions on sphere acquire at least one pole. A
single pole at $Y=Y_i$ \be \psi_i(Y) = q_i/(Y-Y_i) \ee produces
the beam in angular directions \be Y_i=e^{i\phi_i} \tan \frac
{\theta_i}{2} \label{Yi} .\ee

The function $\psi(Y)$ acts immediately on the function $H$ which
determines the metric and the position of the horizon.
 The analysis showed, \cite{BurA},
 that electromagnetic beams have very strong back reaction to metric
 and deform topologically the horizon, forming the holes which
allows matter to escape interior (see fig.3).

The exact KS solutions may have arbitrary number of beams in
different angular directions $Y_i=e^{i\phi_i} \tan \frac
{\theta_i}{2}.$ The corresponding function \be \psi (Y) = \sum _i
\frac {q_i} {Y-Y_i}, \label{psiY}\ee leads to the horizon with
many holes.
 In the far zone the beams tend to the
known exact singular pp-wave solutions. The  considered in
\cite{Multiks} multi-center KS solutions showed that the beams are
extended up to the other matter sources, which may also be assumed
at infinity.

The stationary KS beamlike solutions may be generalized to the
time-dependent wave pulses, \cite{BurAxi}, which tend to exact
 solutions in the low-frequency limit.

Since the horizon is extra sensitive to electromagnetic
excitations, it may also be sensitive to the vacuum
electromagnetic field which is exhibited classically as a Casimir
effect, and it was proposed in \cite{BurExa} that the vacuum beam
pulses shall produce a fine-grained structure of fluctuating
microholes in the horizon, allowing radiation to escape interior
of black-hole, as it is depicted on Fig.3.

The function $\psi(Y,\t) ,$ corresponding to beam pulses, has to
depend on retarded time $\t $ and satisfy to the obtained in
\cite{DKS} nonstationary Debney-Kerr-Schild (DKS) equations
leading to the extra long-range radiative term $\gamma(Y,\t)Z .$
The expression for the null electromagnetic radiation take the
form, \cite{DKS}, $F^\mn = Re \cF _{31} e^{3\m}\wedge e^{1\n},$
where \be  \cF _{31}=\gamma Z - (AZ),_1 \ , \ee  $Z=P/(r+ia\cos
\theta), \quad P=2^{-1/2}(1+\Y\bar \Y) ,$ and the null tetrad
vectors have the form $e^{3\m}= Pk^\m, \ e^{1\m} = \d_\Y e^{3\m}.$

The long-term attack on the DKS equations has led to the obtained
in \cite{BurExa} time-dependent solutions which revealed a
holographic structure of the fluctuating Kerr-Schild spacetimes
and showed explicitly that the electromagnetic radiation from a
black-hole interacting with vacuum contains two components:

a) a set of the singular beam pulses (determined by function
$\psi(Y,\t) ,$) propagating along the Kerr PNC and breaking the
topology and stability of the horizon;

b) the regularized radiative component (determined by
$\gamma_{reg}(Y,\t)$) which is smooth and,  similar to that of the
the Vaidya `shining star' solution, determines evaporation of the
black-hole, \be\dot m = - \frac 12 P^2<\gamma_{reg}\bar
\gamma_{reg}> .\ee

The mysterious twosheetedness of the KS geometry plays principal
role in the holographic black-hole spacetime \cite{BurExa},
allowing one to consider action of the electromagnetic in-going
vacuum as a time-dependent process of scattering. The obtained
solutions describe excitations of electromagnetic beams on the KS
background, the fine-grained fluctuations of the black-hole
horizon, and the consistent back reaction of the beams to metric
\cite{BurA,BurExa}. The holographic space-time is twosheeted and
forms a fluctuating pre-geometry which reflects the dynamics of
the singular beam pulses. This pre-geometry is classical, but  has
to be still regularized to get the usual smooth classical
space-time. In this sense, it takes an intermediate position
between the classical and quantum gravity.

Meanwhile, the function $Z =P/(r+ia \cos\theta)$ tends to infinity
 near the Kerr ring, indicating that any electromagnetic
excitation of the KN geometry should generates the related
singular traveling waves along the Kerr singular ring, and
therefore, the `axial' singular beams turn out to be topologically
coupled with the `circular' traveling waves, see Fig.4. The both
these excitations travel at the speed of light, and the `axial'
beams tend asymptotically (by $r\to \infty$) to the  pp-wave
(plane fronted wave) solutions, for which the vector $k_\m$ in
(\ref{ksm}) forms a covariantly constant Killing direction.

\begin{figure}[h]
\begin{minipage}{17pc}
\includegraphics[width=15pc]{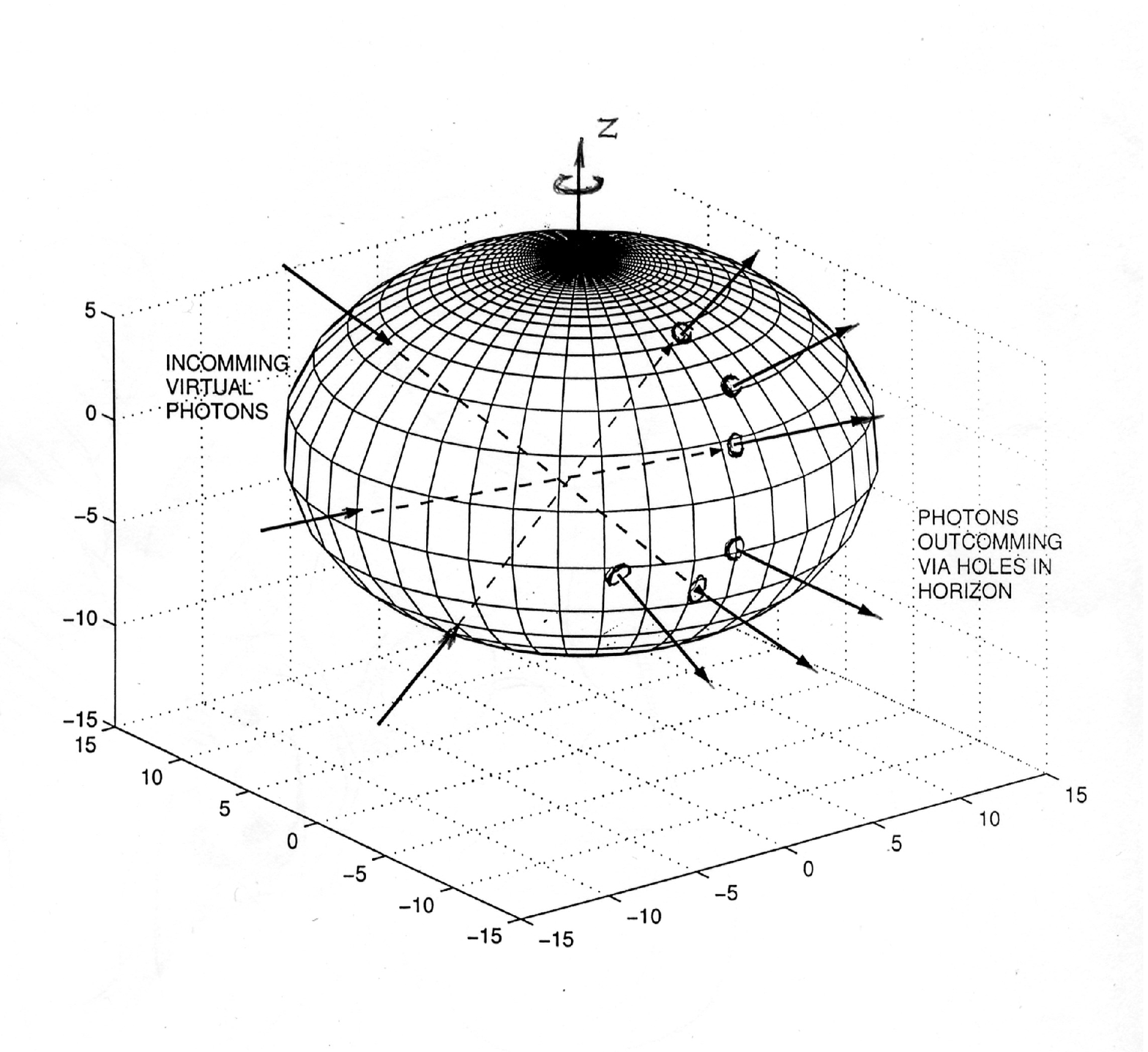}
 \caption{Excitations of a black hole by a
 weak EM field creates a series of fluctuating twistor-beams (outgoing pp-waves) which
 perforate the black hole horizon, covering it by the fluctuating micro-holes.}
\end{minipage}\hspace{6pc}%
\begin{minipage}{14pc}
\includegraphics[width=17pc]{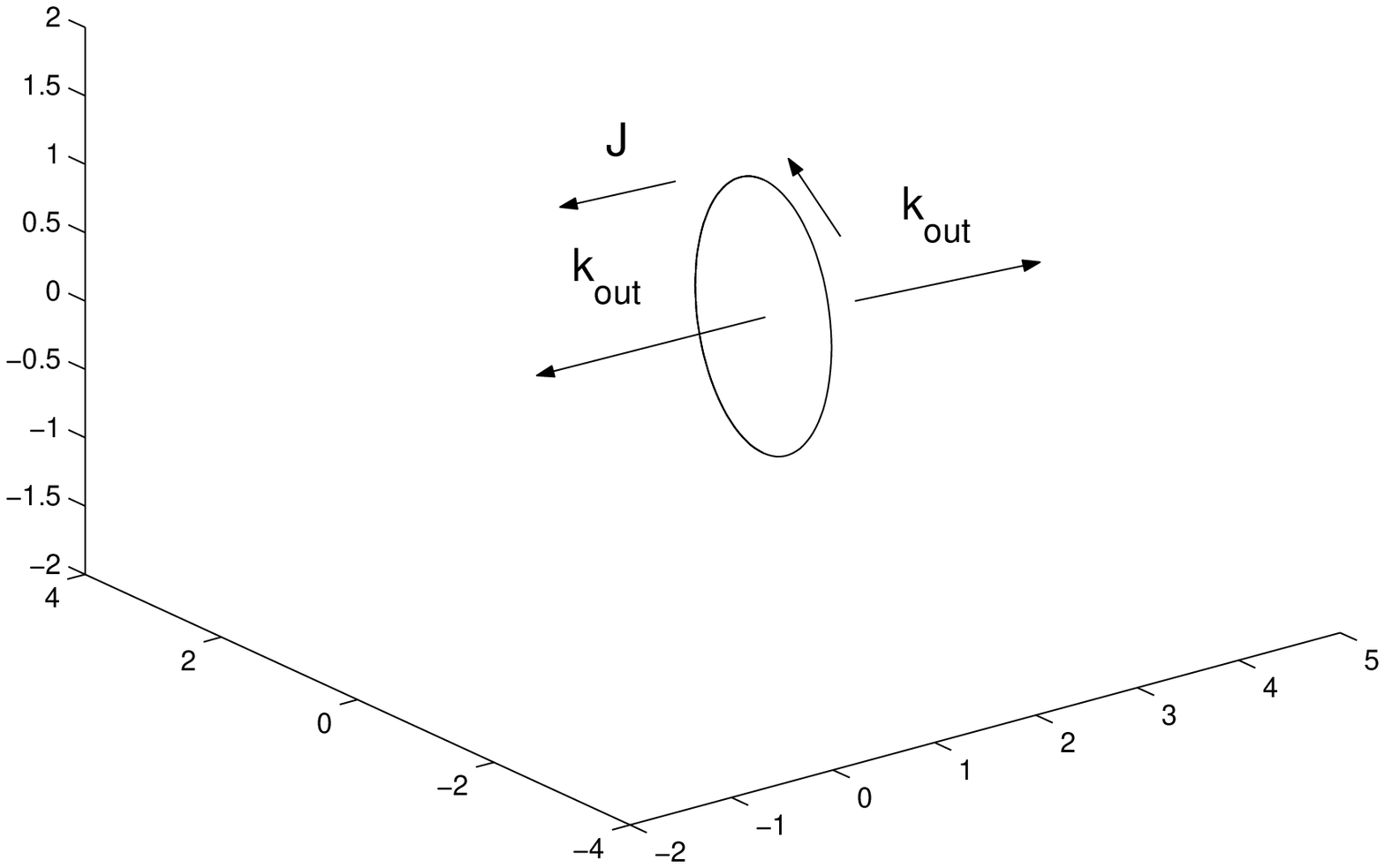}
\caption{Skeleton of the Kerr geometry \cite{BurAxi,BurExa} formed
by the topologically coupled `circular' and `axial' strings.}
\end{minipage}
\end{figure}

 The pp-waves take very important role in superstring theory, forming
 the singular classical solutions
to the low-energy string theory \cite{HorSt}. The string solutions
are compactified to four dimensions and the singular pp-waves are
regarded as the massless fields around a lightlike fundamental
string. It is suspected that the singular source of the string
will be smoothed out in the full string theory, taking into
account all orders in $\alpha '.$  In the nonperturbative approach
based on analogues between the strings and solitons, the pp-wave
solutions are considered as fundamental strings
\cite{DabhGibHarvRR}. The pp-waves may carry traveling
electromagnetic and gravitational waves which represents
propagating modes of the fundamental string \cite{Garf}. In
particular, the generalized pp-waves represent the singular
strings with traveling electromagnetic waves \cite{BurAxi,Tseyt}.
It has been noticed that the field structure of the Kerr singular
ring is similar to a closed pp-wave string \cite{Bur0,IvBur}. This
similarity
 is not incidental, since many solutions to the Einstein-Maxwell theory
turn out to be particular solutions to the low energy string
theory with a zero (or constant) axion and dilaton fields. Indeed,
 the bosonic part of the action for the low-energy string theory takes
 after compactification to four dimensions the following form, \cite{ShapTriWilc},
 \be S = \int  d^4 x \sqrt {-g} ( R -2 (\d \phi)^2 - e^{-2\phi} F^2
 -  \frac 12 e^{4\phi}(\d a)^2 - a F\tilde F) \label{Seff}, \ee
which contains the usual Einstein term
 $S_g = \int  d^4 x \sqrt {-g} R $ completed by the kinetic term for dilaton field
 $-2 (\d \phi)^2$ and by the scaled by $\phi$ electromagnetic
 field. The last two terms are related with axion field $a$ and
 represent its nonlinear coupling with dilaton field $- \frac 12 e^{4\phi}(\d a)^2$
 and interaction of the axion with the dual electromagnetic field
$ \tilde F_\mn =
 \epsilon_\mn ^{\lambda \rho} F_{\lambda \rho}.$

It follows immediately that {\it any solution of the Einstein
gravity, and in particular the Kerr solution, is to be exact
solution of the effective low energy string theory} with a zero
(or constant) axion and dilaton fields. Situation turns out to be
more intricate for the Einstein-Maxwell solutions since the
electromagnetic invariant $F^2$  plays the role of the source of
dilaton field. Similarly, the term $F\tilde F$ turns out to be the
source of the axion field. The stringy analog to the Kerr-Newman
solution with nontrivial axion and dilaton fields was obtained by
Sen \cite{Sen}, and it was shown in \cite{BurStr}, that the field
around the singular string in the `axidilatonic' Kerr-Sen solution
is very similar to the field around a heterotic
string.\fn{Peculiarity of the heterotic strings is related with
the lightlike current and the lightlike bosonic traveling modes of
one direction.} This proximity of the pure gravitational strings
\cite{IvBur} to the low-energy string theory allows us to consider
the Kerr singular ring as a closed heterotic string and the
corresponding traveling waves as its lightlike propagating modes.
The axidilaton field is related with the string tension, and
therefore, the nontrivial solutions to the low energy string
theory  should be very important, allowing one to estimate the
mass-energy of the excited string states. The structure of the
Lagrangian (\ref{Seff}) shows that the axion field involves the
dual magnetic field, and therefore, the complex axidilaton
combination may generate  the duality rotation and create an
additional twist of the electromagnetic traveling waves. However,
the exact solutions of this type are so far unknown. Note also
that the axidilaton field appears naturally in the based on the
special-K\"ahler geometry 4D models of black holes in
supergravity, which may have important consequences for the models
of regularized KN solution \cite{BurSol}.

Assuming that the lightlike string forms a core of the electron
structure, we have to obtain a bridge to the one-particle quantum
theory.  Traveling waves along the KN closed string generate the
spin and mass of the stringlike particle. Physically, it is
equivalent to the original Wheeler's model of `mass without mass'
\cite{Wheel}. In the next section we  show  emergence of the Dirac
equation from this physical picture.

\section{Mass without mass}  The puzzle of "zitterbewegung"  and the
 known processes of
 annihilation of the electron-positron pairs brought author in 1971 to
the Wheeler "geon" model of the "mass without mass" \cite{Wheel}.
 In \cite{BurGeon0} we considered a massless particle circulating
around z-axis. Its local 4-momentum is lightlike,  \be p_x^2 +
p_y^2 + p_z^2 = E^2 \label{Ephot} ,\ee while the effective
mass-energy was created by an averaged orbital motion,
 \be <p_x^2> +<p_y^2> = \tilde m^2
\label{mPxy} .\ee Averaging (\ref{Ephot}) under the condition
(\ref{mPxy}) yields  \be <p_x^2 + p_y^2 + p_z^2> = \tilde m^2
+p_z^2 = E^2 \label{mPzE} .\ee Quantum analog of this model
corresponds to a wave function $\psi(\vec x,t) $ and operators, $
\vec p \to \hat {\vec p} = -i\hbar \nabla , \quad \hat E= i \hbar
\d_t .$  From (\ref{Ephot}) and (\ref{mPxy}) we obtain the
D'Alembert equation $\d^\m \d_\m \psi =0$ and the constraint
$(\d_x^2 + \d_y^2)\psi=0 ,$ which for the chosen coordinate system
are reduced to the equations \be (\d_x^2 + \d_y^2)\psi = \tilde
m^2 \psi = (\d_t ^2
 - \d_z^2)\psi \label{msep} ,\ee and may be separated by the ansatz
 \be\psi ={\cal{M}}(x,y)\Psi_0 (z,t)
\label{ans}. \ee  The RHS of (\ref{msep}) yields the usual
equation for a massive particle, $ (\d_t ^2 - \d_z^2)\Psi_0
=\tilde m^2 \Psi_0 ,$ and the corresponding (de Broglie) plane
wave solution \be \Psi_0 (z,t) =\exp{\frac i \hbar (z p_z -Et)} ,
\label{deBr} \ee while the l.h.s. determines the ``internal''
structure factor \be {\cal{M}}_\n={\cal{H}}_\n (\frac {\tilde m}
\hbar \rho)
 \exp \{i\n \phi \} \label{MHan}, \ee
in polar coordinates $\rho, \phi ,$ where
  ${\cal{H}}_\n ( \frac {\tilde m } \hbar \rho) $ are the Hankel
functions of index $\n.$   ${\cal{M}}_\n$ are eigenfunctions of
operator
 $\hat J_z = \frac \hbar i \d_\phi $ with eigenvalues $J_z=
\n\hbar .$ For electron we have $J_z= \pm \hbar/2, \quad \n=\pm
1/2 ,$ and the factor \be {\cal{M}}_{\pm 1/2}= \rho^{-1/2}\exp \{
i (\frac {\tilde m} \hbar \rho \pm \frac 12 \phi )\} \label{M12}
\ee creates a singular ray along $z$-axis, which forms a branch
line, and the wave function is twovalued.

 There are diverse generalizations of this solution. First of all,
 there may be obtained the corresponding wave functions based on the
 eigenfunctions of the operator of the total angular momentum and
 simultaneously of the spin projection operator. Next, the treatment
 may be considered in a Lorentz covariant form for arbitrarily
positioned and oriented wave functions. And finally, the
corresponding spinor models, together with  all the corresponding
spinor solutions may also be obtained (see \cite{Beyond}).

Principal peculiarity of the obtained massless model is that the
usual plane wave functions are replaced by the vortex waves
generating the spin and mass of the particle-like solutions of the
{\it massless} equations, and therefore, the contradiction between
the massive wave equation and the lightlike zitterbewegung
(determined by the Dirac operators $\alpha$) disappears. On the
other hand, the wave functions (\ref{ans}) are factorized into the
usual plane waves $\Psi_0 (z,t)$ and the  string-like singular
factors ${\cal{M}}(x,y)$ playing the role of singular carriers of
de Broglie waves, which reproduces de Broglie's wave-pilot
conception, which is however principally different from the
corresponding Bohm model.

It should also be mentioned that the characteristic spinor
twovaluedness appears also for scalar waves, as a consequence of
the topological twosheetedness generated by the singular branch
line. In the Kerr geometry this `axial' branch line is linked with
the Kerr `circular' branch line (Figure 1.), forming a
topologically nontrivial spacetime structure of the KN geometry,
(Figure 2.).

\section{Regularization: Electron as a gravitating soliton}

The experimentally indicated KN background of the electron
exhibits the closed singular string which contradicts to the
Quantum assumptions that the gravity is negligible and the
background  is flat. As a result, the justification of the Dirac
electron theory and  QED requires {\it regularization of the KN
metric,} which should be performed with invariability of its
asymptotic form. Similar regularization of the singular strings of
the low-energy string theory is assumed in the full string theory
\cite{BBS}. For the KN solution this problem is close related with
general problem of the regularization of the black hole
singularity \cite{Dym0} and with the old problem of the regular
source of the KN solution \cite{BurBag,BEHM,GG}.

\subsection{Gravitational aspect} The used by Israel truncation of the negative sheet of
the KN solution \cite{Isr} led to the disklike model of the KN
source which retained the Kerr singular ring. It was replaced by
L\'opez by the regular model of a rigidly rotating charged bubble
with a flat interior \cite{Lop}, which was a prototype of the
gravitating soliton model \cite{BurSol}. The singular region of
the KN solution is rejected in the L\'opez model and replaced by
the flat space-time, forming a bubble with flat interior. One
should retain the asymptotic form of the external KN solution and
provide a smooth matching of the external metric with the flat
bubble interior. It is achieved by the special chose of the bubble
boundary $r_{b}$ which is determined by the condition $H(r_{b})=0
.$ From (\ref{H}) one obtains \be r_b =r_e = e^2/(2m) ,\label{rb}
\ee where $r_b$ is the Kerr ellipsoidal radial coordinate,
(\ref{oblate}). As a result, the regular KN source takes the form
of an oblate disk of the Compton radius $r_c \approx a
=\hbar/(2m)$ with the thickness $r_e = e^2/(2m) $ corresponding to
the known 'classical size' of the electron. One sees that the
consistent regularization needs an extension up to the Compton
distance.  The electromagnetic field is also regularized, and the
L\'opez bubble model represents a charged and rotating singular
shell.
 The corresponding smooth and regular rotating sources of the
 Kerr-Schild class were considered in \cite{BurBag,BEHM} on the base
 of the generalized  KS class of metrics suggested by G\"rses and G\"ursey
 in \cite{GG}. The function $H $ in the generalized KS form of
 metric
\[g_\mn = \eta_\mn + 2 H
k_\m k_\n ,\] is taken to be \be H=f(r)/(r^2 + a^2 \cos ^2\theta)
, \label{HGG}\ee where the function $f(r)$ interpolates between
the inner regular metric and the external KN solution. By such a
deformation, the Kerr congruence, determined by the vector field
$k^\m( m) \in M^4 ,$ should retain the usual KS form (\ref{kY}).

It  allows one to suppress the Kerr singular ring
 ($r=\cos \theta =0$) by a special choice of the function $f(r).$

The regularized solutions have tree regions:

i) the Kerr-Newman exterior, $r>r_0 $, where $f(r)=mr -e^2/2,$

ii) interior $r<r_0-\delta $, where $f(r) =f_{int}$ and function
$f_{int}=\alpha r^n ,$ and $n\ge 4$ to suppress the singularity at
$r=0,$ and provide the smoothness of the metric up to the second
derivatives.

iii) intermediate region providing
 a smooth  interpolation between i) and ii).

\subsection{Material aspect}

To remove the Kerr-Newman singularity, one has to set for the
internal region
\[f_{int}=\alpha r^4 .\]
In this case, the Kerr singularity is replaced by a regular
rotating internal space-time with a constant curvature, $ R=-24
\alpha $ \cite{BurBag,BEHM}.

 The functions
\begin{equation}
D= - \frac{f^{\prime\prime}} {\Sigma}, \quad G=
\frac{f'r-f}{\Sigma^2} \label{Gt}.
\end{equation}
determine stress-energy tensor  in the orthonormal tetrad
$\{u,l,m,n\}$  connected with the Boyer-Lindquist coordinates,
\begin{equation}
T_{ik} = (8\pi)^{-1} [(D+2G) g_{ik} - (D+4G) (l_i l_k -  u_i
u_k)]. \label{Tt}
\end{equation}
In the above formula, $u^i$ is a timelike vector field given by
$$
u^i=\frac 1{\sqrt{\Delta\Sigma}}(r^2+a^2,0,0,a) .
$$
This expression shows that the matter of the source is separated
into ellipsoidal layers corresponding to constant values of the
coordinate $r$, each layer rotates with angular velocity
$\omega(r)= \frac {u^{\phi}}{u^0}=a/(a^2+r^2)$. This rotation
becomes rigid only in the thin shell approximation $r=r_0$. The
linear velocity of the matter w.r.t. the auxiliary Minkowski space
is $v=\frac {a \sin \theta}{\sqrt {a^2 + r^2}}$, so that on the
equatorial plane $\theta =\pi /2$, for small values of $r$ ($r\ll
a $), one has $v \approx c=1$, that corresponds to an oblate,
relativistically rotating disk.

The energy density $\rho$ of the material satisfies to
$T^i_ku^k=-\rho u^i$ and is, therefore,  given by
\begin{equation}
\rho = \frac{1}{8\pi} 2G. \label{rhot}
\end{equation}
Two distinct spacelike eigenvalues, corresponding to the radial
and tangential pressures of the non rotating case are
\begin{equation}
p_{rad} = -\frac{1}{8\pi} 2G=-\rho, \label{pradt}
\end{equation}
\begin{equation}
p_{tan} = \frac{1}{8\pi}(D+ 2G)=\rho +\frac{D}{8\pi}. \label{prtt}
\end{equation}
In the exterior region function $f$ must coincide with Kerr-Newman
solution, $f_{KN} = mr -e^2/2$.

There appears a transition region
 placed in between the boundary of the matter object  and the de Sitter core.
  This transition region has to
be described by a smooth function $f(r)$ which interpolates
between the functions $f_{int}(r)$ and $f_{KN}(r)$.  Graphical
analysis allows one to determine position  of the bubble boundary
$r_b ,$ \cite{BEHM,RenGra}. For the L\'opez model of the flat
interior $f_{int}=0$ and $r_b =r_e =e^2/(2m).$ The case $\alpha
>0$ corresponds to de Sitter interior and uncharged source.
 There is only one intersection between $f_{int}(r)=\alpha r^4$ and
$f_{KN}(r)=mr$. The position of the transition layer will be $r_b
=(m/\alpha)^{-1/3}$. The second derivative of the corresponding
interpolating function will be negative at this point, yielding an
extra contribution to the positive tangential pressure in the
transition region.

\subsection{Chiral field model and the Higgs field} In accordance with the Einstein equations, the
considered smooth and regular metric should be generated by a
system of the matter fields forming a classical source of the
vacuum bubble. In the suggested in \cite{BurSol} soliton model,
the smooth phase transition  from the external KN solution to the
internal `pseudovacuum state'  is generated by a supersymmetric
set of chiral fields $\Phi^i, \quad i=1,2,3 \ ,$
\cite{BurSol,BurBag,BurCas} controlled by the suggested by Morris
\cite{Mor} super-potential \be W= \lambda Z(\Sigma \bar \Sigma
-\eta^2) + (cZ+ \m) \Phi \bar \Phi ,\ee where $c, \ \m, \ \eta, \
\lambda$ are the real constants, and we have set $\Phi^1 =\Phi, \
\Phi^2=Z $ and $\Phi^3 =\Sigma .$ The potential is determined by
the usual relations of the supersymmetric field theory
\cite{WesBag} \be V(r)=\sum _i |\d_i W|^2 ,\ee where $ \d_1 =
\d_\Phi , \ \d_2 = \d_Z , \ \d_3 = \d_\Sigma .$ The vacuum states
are determined by the conditions $\d_i W =0 $ which yield $V=0$

i) for `false' vacuum ($r<r_0$): $Z=- \m/c; \Sigma=0; |\Phi|=
\eta\sqrt{\lambda/c},$ and also

ii) for `true' vacuum ($r>r_0$) : $ Z=0; \Phi=0; \Sigma=\eta .$

\noindent which provides a phase transition from the external KN
`vacuum state', $ V_{ext}=0 ,$  to a flat internal `pseudovacuum'
state, $ V_{int}=0 ,$ providing regularization of the Kerr
singular ring, \cite{BurBag,BurCas,BurSol}. One of the chiral
fields, $\Phi^1 ,$ is set as the Higgs field $\Phi^1 \equiv \Phi =
\Phi_0 \exp(i\chi ).$ As a result of the phase transition, the
Higgs field $\Phi_0 \exp(i\chi )$ with a nonzero vev $\Phi_0$ and
the phase $\chi$ fills interior of the bubble and regularizes the
electromagnetic Kerr-Newman field by the Higgs mechanism of broken
symmetry.
\subsection{Regularization of the electromagnetic KN field}
The electromagnetic KN field inside the bubble interacts with the
Higgs field in agreement with  Landau-Ginzburg type field model
 with the Lagragian \be {\cal L}_{NO}= -\frac 14 F_\mn F^\mn + \frac 12
(\cD_\m \Phi)(\cD^\m \Phi)^* + V(r), \label{LNO}\ee where $ \cD_\m
= \nabla_\m +ie \alpha_\m $  are to be covariant derivatives. This
model was used by  Nielsen and Olesen \cite{NO} for the obtaining
the string-like solutions in superconductivity. The model
determines the current \be I_\m = \frac 12 e |\Phi|^2 (\chi,_\m +
e \alpha_\m) \ee as a source of the Maxwell equations $F^{\n\m}_{
\ \ ;\n}=I^\m .$ This current should vanish inside the bubble,
which sets a relation between incursion of the phase  of the Higgs
field and the value of the vector potential of the KN solution on
the boundary of the bubble \be \chi,_\m =- e \alpha_\m^{(str)} .
\label{Iinside}\ee  The maximal value of the regularized vector
potential $\alpha_\m^{(str)}$  is reached on the boundary of the
bubble. In the agreement with (\ref{rb}) and (\ref{ksGA}) we have
the vector relation \be \alpha_\m^{(str)} = \alpha_\m (r_{b}) =
2m/e , \label{Amax}\ee which results in two very essential
consequences \cite{BurSol}:
\begin{itemize}
\item  the Higgs field (matched with the regularized KN
electromagnetic field) forms a coherent vacuum state oscillating
with the frequency $\omega=2m ,$ which is a typical feature the
``oscillon'' soliton models.

\item  the regularized KN electromagnetic potential
$\alpha^{(str)}_\m$ forms on the boundary of the bubble a closed
Aharonov-Bohm-Wilson quantum loop $ \oint e\alpha^{(str)}_\phi
d\phi=-4\pi ma \label{WL} ,$ which determines quantized  spin of
the soliton, $J=ma=n/2, \ n=1,2,3,...$
\end{itemize}

Does the KN model of electron contradict to Quantum Theory? It
seems ``yes'', if one speaks on the "bare" electron. However, in
accordance with QED, vacuum polarization creates in the Compton
region a cloud of virtual particles forming a "dressed" electron.
This region gives contribution to electron spin, and performs a
procedure of renormalization, which determines physical values of
the electron charge and mass. Therefore, speaking on the
``dressed'' electron, one can say that the real contradiction
between the KN model and the Quantum electron is absent.

 Note that dynamics of the virtual particles in QED is
chaotic and can be conventionally separated from the
``bare''electron. In the same time,
 the vacuum state inside the Kerr-Newman soliton forms a {\it coherent
oscillating state} joined with a closed Kerr string. It represents
an {\it integral whole of the extended electron,} its `internal'
structure which cannot be separated from a ``bare'' particle. In
any case, the Kerr string appears as an analogue of the pointlike
bare electron.

\section{Conclusion}
We have showed that gravity definitely indicates presence of a
closed string of the Compton radius $a=\hbar/(2m)$ in the electron
background geometry. This string has gravitational origin and is
close related with the fundamental closed strings of the low
energy string theory. Corpuscular aspect of the  traveling waves
along the Kerr string allows us to `derive'  the Dirac equation.
The original Dirac theory is modified in this case: the wave
functions are factorized and acquire the singular stringlike
carriers. As a result, the new wave functions turn out to be
propagating along the `axial' singular strings, which is
reminiscent of the de Broglie wave-pilot conjecture. Therefore,
the gravitational KN closed string represents a bridge between
gravity, superstring theory and the Dirac quantum theory towards
the consistency of these theories. We arrive at the extremely
unexpected conclusion that Gravity, as a basic part of the
superstring theory, may lie beyond Quantum theory and play a
fundamental role in its `emergence'.

The observable parameters of the electron determine unambiguously
the Compton size of the Kerr string. This size is very big with
respect to the modern scale of the experimental resolutions, and
it seems, that this string should be experimentally detected.
However, the high-energy scattering detects the pointlike electron
structure down to $10^{-16} cm $. One of the explanations of this
fact, given in \cite{BurTwi}, is related with the assumption that
interaction of the KN particles occurs via the lightlike KN
`axial' strings. Just such a type of the `direct' lightlike
interaction follows from the analysis of the Kerr theorem for
multiparticle Kerr-Schild solutions, \cite{Multiks}. So far as the
KN circular string is also lightlike, the lightlike photon can
contact it only at one point. The resulting scattering of the Kerr
string by the \emph{real} photons of high energy can exhibit only
the pointlike interaction, and neither form of the string, nor its
extension cannot be recognized. To recognize the shape of the
string as a whole, it is necessary two extra conditions:

a) a {\it relative low-energy} resonance scattering with the
wavelengths comparable with extension of the string. It means that
there must be a scattering with a low-energy momentum transfer,
i.e. with a small Bjorken parameter $x= q_t /P .$

b) simultaneously, to avoid the pointlike contact interaction, the
scattering should be deeply virtual, i.e. the square of the
transverse four-momentum transfer should satisfy $Q^2 >> m^2.$

Both these conditions correspond to the novel tool -- the Deeply
Virtual Compton Scattering (DVCS) described by the theory of
Generalized Parton Distribution (GPD) \cite{Rad,Ji}, which allows
one to probe the shape of the elementary particles by the
``non-forward Compton scattering'' \cite{Hoyer}. If the predicted
Kerr-Newman string will be experimentally recognized in the core
of electron structure, it could be great step in understanding
Quantum theory towards to Quantum Gravity.

\subsection{Acknowledgments} Author thanks M. Bordag, A. Efremov,
D. Stevens, O. Teryaev and F. Winterberg for conversations and
useful references, and also Yu. Danoyan, A. Efremov,  D. Gal'tsov,
A. Khrennikov, T. Nieuwenhuizen, Yu. Obukhov, O.Selyugin, K.
Stepaniants and G. `t Hooft for useful discussions, and also
especially thankful to A. Radyushkin for discussions concerning
the GPD application.


\section*{References}
\begin{thebibliography}{9}

\bibitem{HorSt}  Horowitz G and  Steif A 1990 Spacetime
Singularities in String Theory {\it Phys.\ Rev.\ Lett.}  {\bf 64}
260

\bibitem{Tseyt} Tseytlin A 1995 Exact solutions of closed string
theory {\it Class. Quant. Grav.} {\bf 12} 2365
[arXiv:hep-th/9505052]

\bibitem{Coley}  Coley A A  2002 A Class of Exact Classical Solutions to String Theory {\it Phys. Rev. Lett.}  {\bf
89} 281601

\bibitem{Car} Carter B 1968 Global structure of the
Kerr family of gravitational fields {\it
  Phys.\ Rev.} {\bf 174} 1559

\bibitem{DKS}  Debney G C, Kerr R P and Schild A 1969
  Solutions of the Einstein and Einstein-Maxwell equations {\it
  J.\ Math.\ Phys.}  {\bf 10} 1842


\bibitem{BurSol} Burinskii A 2010 Regularized Kerr-Newman solution as a
gravitating soliton, {\it J. Phys. A: Math. Theor.} {\bf 43}
392001 [arXiv: 1003.2928].



\bibitem{BurTwi} Burinskii A 2004 Twistorial analyticity and three stringy
systems of the Kerr spinning  particle {\it Phys.\ Rev. }  D {\bf
70} 086006 [arXiv:hep-th/0406063]

\bibitem{BurKN}
Burinskii A The Dirac-Kerr-Newman electron 2008 {\it  Grav.\
Cosmol.} {\bf 14} 109
  [arXiv:hep-th/0507109]


\bibitem{Isr}
Israel W 1970 Source of the Kerr metric {\it  Phys.\ Rev.}  D {\bf
2} 641


\bibitem{BurGeonIII}   Burinskii A Ya 1972 Microgeon with Kerr metric
{\it Abstracts of the III Soviet Gravitational Conference} (
Yerevan, 1972, p.217, in Russian).

\bibitem{Bur0}   Burinskii A Ya 1974 Microgeons with spins {\it
Sov.\ Phys.\ JETP } {\bf 39} 193

\bibitem{IvBur}  Ivanenko D D and Burinskii A Ya 1975
Gravitational strings in the models of elementary particles {\it
Izv.\ Vuz.\ Fiz.}  {\bf 5} 135

\bibitem{Lop} L\'opez C A 1984 An extended model of the electron in general relativity
  {\it Phys.\ Rev.}  D {\bf 30} 313


\bibitem{BurSen} Burinskii A 1995
Some properties of the Kerr solution to low-energy string theory
{\it  Phys. Rev.}  D {\bf 52}, 5826 [arXiv:hep-th/9504139]


\bibitem{BurStr} Burinskii A Ya 1994
String-like structures in complex Kerr geometry In: {\it
Relativity Today }  (Edited by R.P.Kerr and Z.Perj\'es,
Akad\'emiai Kiad\'o, Budapest) p.149  [arXiv:gr-qc/9303003]


\bibitem{BurBag} Burinskii A 2002 Supersymmetric Superconducting Bag as a Core
of Kerr Spinning Particle {\it Grav. Cosmol.} \textbf{8} 261
[arXiv:hep-th/0008129]


\bibitem{Arc}
 Arcos H I and Pereira J N 2004 Kerr-Newman solution as a Dirac particle
{\it  Gen.\ Rel.\ Grav.}  {\bf 36} 2441
  [arXiv:hep-th/0210103]


\bibitem{Dym} Dymnikova I 2006 Spinning superconducting electrovacuum soliton
 {\it Phys.\ Lett.}  B {\bf 639} 368


\bibitem{TN}  Nieuwenhuizen Th M 2007 The Electron and the Neutrino as Solitons
in Classical Electrodynamics In: {\it Beyond the Quantum} (eds.
Th.M. Nieuwenhuizen et al., World Scientific, Singapure)
pp.332-342


\bibitem{FWil} Wilczek F 2008 {\it The Lightness Of Being} (Basic Books)

\bibitem{LSuss} Susskind L 2008 {\it The Black Hole War} (Hachette Book Group
US)

\bibitem{BurExa} Burinskii A 2010 Fluctuating twistor-beam solutions
and holographic pre-quantum Kerr-Schild geometry {\it J. Phys.:
Conf. Ser.} {\bf 222} 012044  [arXiv:1001.0332]

\bibitem{BurPreQ} Burinskii A 2010 Twistor-Beam Excitations of Black-Holes
and Prequantum Kerr-Schild Geometry {\it Theor. Math.Phys.} {\bf
163}(3), 782  [arXiv:0911.5499]


\bibitem{BjoDr}   Bjorken J and Drell S 1964 {\it Relativistic Quantum
Mechanics} (McGraw Hill Book)


\bibitem{Rad}  Radyushkin A V 1997 Nonforward parton
distributions {\it Phys. Rev.} {\bf D 56} 5524 [hep-ph/9704207]


\bibitem{Ji} Ji X 1977 Gauge-invariant decomposition of nucleon spin
{\it Phys. Rev. Lett.} {\bf 78 }, 610  [hep-ph/9603249].

\bibitem{Hoyer} Hoyer P and Samu Kurki Samu 2010  Transverse shape of the
electron {\it Phys. Rev.} {\bf D 81} 013002 [arXiv:0911.3011].

\bibitem{Ham} Hamity V 1976 An "interior" of the Kerr metric
{\it Phys. Lett.} A {\bf 56} 77



\bibitem{BurAxi}
Burinskii A 2004 Axial stringy system of the Kerr spinning
particle {\it Grav. Cosmol.} {\bf 10} 50  [arXiv:hep-th/0403212]



\bibitem{Wheel} Wheeler J A 1962 {\it Geometrodynamics} (Academic
Press, New York)


\bibitem{BurGeon0}   Burinskii A Ya 1971 On  the particlelike solutions of
the massles wave equations. {\it Abstracts of  the VIII All-Union
Conference on Elementary Particle Theory} (Uzhgorod, January 1971
pp.96-98 in Russian)

\bibitem{Beyond} Burinskii A 2010 Gravitational strings beyond quantum theory: Electron as a closed
string [arXiv:1109.3872]

\bibitem{BurA} Burinskii A 2009 First Award Essay of GRF:
   Instability of black hole horizon with respect to electromagnetic
  excitations
  {\it Gen.\ Rel.\ Grav.} {\bf 41} 2281
  [arXiv:0903.3162 [gr-qc]].

\bibitem{Multiks}
Burinskii A 2005 Wonderful consequences of the Kerr theorem  {\it
Grav. Cosmology}  {\bf 11} 301  [hep-th/0506006]


\bibitem{DabhGibHarvRR} Dabholkar A, Gibbons G, Harvey J and
Ruiz Ruiz F 1990  Superstrings and Solitons {\it Nucl. Phys.} {\bf
B 340} 33

\bibitem{Garf} Garfinkle D 1992 Black string traveling waves {\it
Phys.\ Rev.} {\bf D  46} 4286




\bibitem{ShapTriWilc}  Schapere A, Trivedi S and Wilczek F 1991
 Dual dilaton dion {\it Modern Phys. Letters} A. {\bf 6} 2677



\bibitem{Sen} Sen A 1992 Rotating charged black hole
solution in heterotic string theory {\it Phys.\ Rev.\ Lett.}  {\bf
69} 1006
\bibitem{BBS} Backer K, Backer M and Schwarz John H 2007 {\it String Theory and M-Theory: A Modern
Introduction} (Cambridge University Press, ISBN 0-521-86069-5)


\bibitem{Dym0} Dymnikova I 1992 Vacuum Nonsingular Black Hole
{\it Gen.Rel.Grav.} {\bf 24} 235

\bibitem{BEHM} Burinskii A, Elizalde E,  Hildebrandt S R and
Magli G 2002 {\it Phys. Rev. D} {\bf 65}, 064039
[arXiv:gr-qc/0109085]

\bibitem{GG}
G\"urses M and G\"ursey F 1975 {\it J. Math. Phys.} {\bf 16} 2385


\bibitem{RenGra} Burinskii A 2006 Renormalization by gravity and the Kerr spinning particle
{\it J. Phys. A: Math. Gen.}  \textbf{39} 6209
[arXiv:gr-qc/0606097]

\bibitem{BurCas} Burinskii A 2002 Casimir Energy and Vacua vor Superconducting Ball in
Supergravity {\it Int. J.Mod. Phys.} {\bf A 17} 920
[arXiv:hep-th/0205127]

\bibitem{Mor}  Morris J R 1996  Supersymmetry and Gauge Invariance Constraints in
a U(1)$\times $U(1)$^{\prime }$-Higgs Superconducting Cosmic
String Model {\it Phys.Rev.} {\bf D53} 2078 [arXiv:hep-ph/9511293]
\bibitem{WesBag} Wess J and Bagger J 1983 {\it Supersymmetry and
Supergravity} (Princeton Univ. Press, Princeton, New Jersey)

\bibitem{NO}
Nielsen H B and Olesen P 1973 {\it Nucl. Phys.} {\bf B61} 45





\end{thebibliography}
\end{document}